# Observation of Lee-Yang zeros

Xinhua Peng[1], Hui Zhou[1], Bo-Bo Wei[2], Jiangyu Cui[1], Jiangfeng Du[1] & Ren-Bao Liu[2]

1. *Hefei National Laboratory for Physical Sciences at Microscale, Department of Modern Physics, and Synergetic Innovation Center of Quantum Information & Quantum Physics, University of Science and Technology of China, Hefei, 230026, China*

2. *Department of Physics, Centre for Quantum Coherence, and Institute of Theoretical Physics, The Chinese University of Hong Kong, Shatin, New Territories, Hong Kong China*

**Lee-Yang zeros are points on the complex plane of magnetic field where the partition function of a spin system is zero and therefore the free energy diverges [1, 2]. Lee-Yang zeros and their generalizations are ubiquitous in many-body systems [3-10] and they fully characterize the analytic properties of the free energy and hence thermodynamics of the systems. Determining the Lee-Yang zeros is not only fundamentally important for conceptual completeness of thermodynamics and statistical physics but also technically useful for studying many-body systems. However, Lee-Yang zeros have never been observed in experiments [11,12], due to the intrinsic difficulty that Lee-Yang zeros would occur only at complex values of magnetic field, which are unphysical. Here we report the first observation of Lee-Yang zeros, by measuring quantum coherence of a probe spin coupled to an Ising-type spin bath [13]. As recently proposed [13], the quantum evolution of the probe spin introduces a complex phase factor, and therefore effectively realizes an imaginary magnetic field on the bath. From the**

**measured Lee-Yang zeros, we reconstructed the free energy of the spin bath and determined its phase transition temperature. This experiment demonstrates quantum coherence probe as a useful approach to studying thermodynamics in the complex plane [14], which may reveal a broad range of new phenomena that would otherwise be inaccessible if physical parameters are restricted to be real numbers.**

After the pioneering works by van der Waals [15,16], Mayer [17,18], and von Hove [19], it had been known that different phases (e.g., liquid and gas phases) of a thermodynamic system have the same microscopic interactions but the free energy of the system encounters a singularity (non-analytic) point in the physical parameter space where the phase transition occurs. A rigorous relation between the analytic properties of free energies and thermodynamics (in particular, phase transitions) was established by Lee and Yang in a seminal paper published in 1952 through continuation of the free energy to the complex plane of physical parameters [1]. Lee and Yang considered a general Ising model with ferromagnetic interaction $J_{ij} > 0$ under a magnetic field $h$, with Hamiltonian $H(h) = -\sum_{i,j} J_{ij} s_i s_j - h \sum_j s_j$, where the spins $s_j$ take values $\pm 1/2$. The partition function of $N$ spins at temperature $T$ (or inverse temperature $\beta \equiv 1/T$) $\Xi(\beta, h) \equiv \sum_{\text{all states}} \exp(-\beta H)$ can be written into an $N$-th order polynomial of $z \equiv \exp(-\beta h)$ as $\Xi = \exp(\beta N h/2) \sum_{n=0}^{N} p_n z^n$, where $\exp(-\beta H)$ is the Boltzmann factor (the probability in a state with energy $H$, up to a normalization factor) and the coefficients $p_n$ can be interpreted as the partition function in zero magnetic field under the constraint that $n$ spins are at state $-1/2$. The free energy $F$ is related to the partition function by $F = -T \ln(\Xi)$. Obviously, the zeros of the partition function (where $\Xi = 0$) are the singularity points of the free energy and hence fully determine the analytic properties of the free energy. If the Lee-Yang zeros are determined, the partition function can be readily reconstructed as $\Xi = p_0 \exp(\beta Nh/2) \prod_{n=1}^{N}(z - z_n)$. Since the Boltzmann factor is always positive for real interaction parameters and real temperature, zeros of the partition function would occur only on the complex plane of the physical parameters. Lee and Yang proved that for the ferromagnetic Ising model the $N$ zeros of the partition function all lie within an arc on the unit circle in the complex plane of $z$ (corresponding to pure imaginary values of the external field) [2]. At sufficiently low temperature ($T \leq T_C$), the end points of the arc, i.e., the Yang-Lee singularity edges [20, 21] approach to the real axis of $h$ at the thermodynamic limit ($N \to \infty$). Thus the free energy encounters a singularity point on the real axis of the magnetic field, which means the onset of a phase transition.

The Lee-Yang zeros exist universally in many-body systems. The Ising models describe a broad range of physical systems, including anistropic magnets, alloys, and lattice gases. The Lee-Yang theorem, first proved for ferromagnetic Ising models of spin-1/2, was later generalized to general ferromagnetic Ising models of arbitrarily high spin [3-5] and to other interesting types of interactions [6-9]. For general many-body systems, the Lee-Yang zeros may not be distributed along a unit circle but otherwise present similar features as in ferromagnetic Ising models. Lee-Yang zeros have also been generalized to zeros of partition functions in the complex plane of other physical parameters (such as Fisher zeros in the complex plane of temperature [10]). The Lee-Yang zeros (or their generalizations) fully characterize the analytic properties of free energies and hence thermodynamics of the systems. Therefore determining the Lee-Yang zeros is not only fundamentally important for a complete picture of thermodynamics and statistical physics (by continuation to the complex plane) but also technically useful for studying thermodynamics of many-body systems.

Experimental observation of Lee-Yang zeros, however, has not been made before. The previous experiments could only indirectly derive the densities of Lee-Yang zeros from susceptibility measurement plus analytic continuation [11,12]. The difficulty is intrinsic: The Lee-Yang zeros would occur only at complex values of external fields or temperature, which are unphysical.

A recent theoretical discovery about the relation between partition functions and probe spin coherence [13] makes it experimentally feasible to access the complex plane of physical parameters. Wei & Liu found that the coherence of a central spin embedded in an Ising-type spin bath is equivalent to the partition function of the Ising bath under a complex magnetic field. The imaginary part of the magnetic field is realized by the time since the quantum coherence of the central spin is a complex phase factor as a function of time. The Lee-Yang zeros of the partition function are one-to-one mapped to the zeros of the central spin coherence, which are directly measurable.

Here we report on experimental observation of Lee-Yang zeros via central spin decoherence measurement, using liquid-state nuclear magnetic resonance (NMR) of trimethylphosphite (TMP) molecules (Fig. 1a) to simulate a coupled probe-bath system. A TMP molecule contains nine equivalent $^1$H nuclear spins ($\mathbf{s}_1, \mathbf{s}_2, \ldots, \mathbf{s}_9$, regarded as the bath in our experiments) and one $^{31}$P nuclear spin ($\mathbf{s}_0$, the probe spin) [22]. In the liquid state, the nine $^1$H spins have long-range Heisenberg interaction with strength $J = 2\pi \times 16.75 \text{ sec}^{-1}$, and the $^{31}$P spin has Ising-type interaction with the nine bath spins with a uniform coupling constant $\lambda = 2\pi \times 10.57 \text{ sec}^{-1}$. The probe-bath Hamiltonian

$$H_{\text{TMP}} = -\nu_H \sum_{j=1}^{9} s_j^z - \nu_P s_0^z + J \sum_{1 \leq i < j \leq 9} \mathbf{s}_i \cdot \mathbf{s}_j + \lambda s_0^z \sum_{j=1}^{9} s_j^z,$$

where $\nu_H = 2\pi \times 400.25 \times 10^6 \text{ sec}^{-1}$ and $\nu_P = 2\pi \times 161.92 \times 10^6 \text{ sec}^{-1}$ are the Larmor frequencies of the $^1$H and $^{31}$P nuclear spins under a magnetic field 9.4 Tesla, respectively. The coupling to the $^1$H nuclear spins splits the NMR resonance of the $^{31}$P nuclear spin into 10 peaks corresponding to the 10 quantized polarizations of the 9 $^1$H spins (Fig. 1b). Note that the microscopic Hamiltonian above is of the anti-ferromagnetic Heisenberg type instead of the ferromagnetic Ising type and the magnetic field was strong. To facilitate observation of the Lee-Yang zeros on the unit circle, we used the quantum simulation method to prepare ensembles of the bath that are described by the effective density matrix (see Methods Summary and Supplementary Information for details)

$$\rho_{\text{eff}} \propto \exp(-\beta_{\text{eff}} H_{\text{eff}}), \tag{1}$$

where the effective Hamiltonian $H_{\text{eff}} = -J \sum_{1 \leq i < j \leq 9} s_i^z s_j^z - h \sum_{1 \leq i \leq 9} s_i^z$ has the form of an ferromagnetic Ising model, $\beta_{\text{eff}} = 1/T_{\text{eff}}$ is the effective inverse temperature, and $h$ is the effective magnetic field. Since the effective Ising Hamiltonian commutes with both the Heisenberg interaction of the spin bath and the probe-bath interaction, the microscopic interactions of the coupled probe-bath system would not affect the simulated ensembles.

We initially prepared the probe spin in a superposition state as $|\Psi(0)\rangle = |\uparrow\rangle + |\downarrow\rangle$ and detected its coherence $L(t) \equiv \langle s_0^x \rangle + i \langle s_0^y \rangle$ as a function of time. The experimental scheme is schematically illustrated in Fig. 1c. The coupling between the probe and the

bath resulted in a local magnetic field $b = -\lambda \sum_{j=1}^{j} s_j^z$, which during the quantum evolution of the probe spin induces a phase factor to the state: $|\Psi(t)\rangle = |\uparrow\rangle + \exp(-ibt)|\downarrow\rangle$. Since the local field $b$ depends on the random configuration of the bath spins [$s_j^z$ takes values randomly from $+1/2$ and $-1/2$, according to the distribution in equation (1)], the phase accumulated by the probe spin became more and more random as the time passed by, which led to the probe spin decoherence. The coherence, as characterized by the spin polarization in the $x$-$y$ plane, is the ensemble average of the phase factor, that is [13],

$$L(t) = \langle e^{-ibt} \rangle = \frac{\text{Tr}\left[\exp(-\beta_{\text{eff}} H_{\text{eff}} - ibt)\right]}{\text{Tr}\left[\exp(-\beta_{\text{eff}} H_{\text{eff}})\right]} = \frac{\Xi(\beta_{\text{eff}}, h + it\lambda/\beta_{\text{eff}})}{\Xi(\beta_{\text{eff}}, h)}. \qquad (2)$$

The probe spin coherence, except for the normalization factor $\Xi(\beta_{\text{eff}}, h)$, is equivalent to the partition function of the spin bath with a complex magnetic field $h + i\lambda t/\beta_{\text{eff}}$. It becomes zero when the evolution time $t$ is such that $z = \exp(-\beta_{\text{eff}} h - i\lambda t)$ equals to a Lee-Yang zero. For the ferromagnetic Ising model, all the Lee-Yang zeros lie on the unit circle of $z$, where $h = 0$. Thus in our experiment we set the effective magnetic field $h$ to be zero.

Figure 2 shows the measured probe spin coherence and the Lee-Yang zeros. For the nine-spin Ising bath, there are 9 zeros $t_n$ for $L(t_n) = 0$ within a period of coherence evolution ($0 \leq t \leq 2\pi/\lambda = 0.0946$ sec), which determine the 9 Lee-Yang zeros $z_n = \exp(-i\lambda t_n)$. At room temperature (the laboratory temperature, at which $T/J \to \infty$), all of the nine Lee-Yang zeros are degenerate at $z_n = -1$ or correspondingly $t_n = \pi/\lambda$, as observed in Figs. 2a & 2d. When the simulated temperature $T_{\text{eff}}$ was comparable to or less than the coupling strength of the bath ($J$), the nine zeros were clearly resolved in the probe spin coherence (Figs. 2b, 2c, 2e & 2f). The measured coherence zeros agree well with the theoretically determined Lee-Yang zeros (see Supplementary Information for methods of calculation).

The Lee-Yang zeros fully determine the partition function or free energy of spin systems, and in turn the thermodynamic properties of the systems. This is fundamentally rooted in the fact that the free energies are analytic functions of the physical parameters, except for the singularity points corresponding to the Lee-Yang zeros. Thus we determined the free energies of the Ising model for various temperatures, by measuring the coherence of just one probe spin. This method is significantly simpler and much more experimentally implementable than the existing proposals [23]. The results are shown in Fig. 3, which compares very well to the theoretical calculation of the free energies of the Ising model.

Phase transitions are intimately connected to the Lee-Yang zeros. At or below the phase transition temperature, the Yang-Lee singularity edges (the Lee-Yang zeros with the smallest imaginary part of the magnetic field) [20, 21] approach to the real axis of the magnetic field in the thermodynamic limit ($N \to \infty$). Our finite spin bath had only nine spins, which is far from the thermodynamic limit. But still the phase transition temperature can be inferred from the fact that below the critical temperature, the Lee-Yang zeros become almost uniformly distributed along the unit circle. The uniform distribution of the Lee-Yang zeros at low temperature ($T_{\text{eff}} \ll J$) was indeed observed in Figs. 2c & 2f. The uniform distribution of Lee-Yang zeros led to periodic oscillation of the probe spin coherence. Such periodic oscillation can be understood from the fact that below the critical temperature the bath spins were mostly in the two degenerate, polarized ground states, which led to interference between probe spin precessions under two opposite local fields. As shown in Fig. 4, below the transition temperature, the measured Yang-Lee edge (the first Lee-Yang zero) is almost constant as a function of temperature. The measured phase transition temperature agrees reasonably well with the theoretical calculation considering the small size of our system (see Supplementary Information for methods of calculation).

Thus we not only directly observed the Lee-Yang zeros in experiments for the first time, which conceptually completes the analytic description of statistical physics and thermodynamics, but also demonstrated the feasibility of using probe spin coherence to determine the thermodynamic properties of the baths and more generally, to access thermodynamics on the complex plane of physical parameters [14].

## METHODS SUMMARY

The sample was a 1:1 by volume solution of trimethylphosphite (TMP) and acetone-$d_6$ in a 5-mm NMR tube. All NMR experiments were performed on a Bruker Avance III 400 MHz (9.4 T) spectrometer at room temperature. The $\pi/2$ hard pulse length was approximately 13 μsec on the hydrogen channel and 17 μsec on the phosphor channel. The measured longitudinal and transverse relaxation times were respectively $T_1 = 5.6$ sec and $T_2^* = 0.65$ sec for $^{31}$P and $T_1 = 8.2$ and $T_2^* = 0.25$ sec for $^1$H. In order to observe the Lee-Yang zeros of the Ising model $H_{\text{eff}}$, we first prepared the ensemble of TMP molecules in the initial state $\rho_{in} = \left(1/2 + \varepsilon_P s_0^x\right) \otimes \rho_{\text{eff}}$, where $\varepsilon_P \approx \beta \nu_P$ is the initial polarization of the $^{31}$P spin at room temperature (the probe spin was initially prepared in a pseudo-pure state corresponding to $|\Psi(0)\rangle = |\uparrow\rangle + |\downarrow\rangle$), and $\rho_{\text{eff}} = \exp(-\beta_{\text{eff}} H_{\text{eff}})/\Xi(\beta_{\text{eff}}, h)$ with $\Xi(\beta_{\text{eff}}, h) = \text{Tr}\left[\exp(-\beta_{\text{eff}} H_{\text{eff}})\right]$ (the bath was in thermal equilibrium for the effective Hamiltonian $H_{\text{eff}}$ and the effective inverse temperature $\beta_{\text{eff}}$). As the probe spin coherence is related to the off-diagonal elements of the density matrix, we focused on the term $\rho_{\text{in}}^{\Delta} = s_0^x \otimes \rho_{\text{eff}}$. We created $\rho_{\text{in}}^{\Delta}$ (up to a trivial strength factor) by the temporal averaging method [24]. The states created were confirmed by partial state tomography [26] and the final fidelity [27]

$$F \equiv \text{Tr}[\rho_{\text{exp}} \rho_{\text{in}}^{\Delta}] \bigg/ \sqrt{\text{Tr}(\rho_{\text{exp}}^2) \text{Tr}\left[(\rho_{\text{in}}^{\Delta})^2\right]}$$ was $\approx 0.99$, where $\rho_{\text{exp}}$ is the experimentally simulated density matrix. We performed five different experiments, in each of which we resonantly irradiated two symmetric transitions of the $^{31}$P spin (corresponding to two peaks with opposite frequency shifts in Fig. 1b) by a Gaussian-shaped line-selective pulse [25] with the length of 550 msec. The probe spin coherence

$$L(t) = \text{Tr}\left[e^{-iH_{\text{TMP}}t} \rho_{\text{in}}^{\Delta} e^{iH_{\text{TMP}}t} \left(s_0^x + is_0^y\right)\right]$$ was measured by the free induction decay (FID) of the $^{31}$P spin in NMR. Then we summed the results from the five sets of experiments with weighting coefficients determined by the corresponding Boltzmann factors. The coherence zeros $t_n$ of $L(t)$ and hence the corresponding Lee-Yang zeros $z_n = \exp(-i\lambda t_n)$ were extracted by fitting these experimental data via a polynomial function (or by interpolation).

**FULL METHODS** and associated references are available in the online Supplementary Information.

**Acknowledgements** This work was supported by Hong Kong Research Grants Council - General Research Fund Project 401413, The Chinese University of Hong Kong Focused Investments Scheme, The National Key Basic Research Program of China (Grant Nos. 2013CB921800 & 2013CB848700), National Natural Science Foundation of China (Grant Nos. 11375167, 11227901 & 91021005), and the Strategic Priority Research Program (B) of The Chinese Academy of Sciences (Grant No. XDB01030400).

**Author Contributions** R.B.L. proposed the experiment. X.P. designed the experiment. X.P., H.Z. and J.C. carried out the experiments. X.P. and H.Z. analyzed experimental data. B.B.W. did the theoretical calculation. J.D. supervised the experimental team. R.B.L. wrote the paper with assistance of X.P., H.Z. & B.B.W. All authors discussed the results and commented on the manuscript.

**Author Information** The authors declare no competing financial interests. Correspondence and requests for materials should be addressed to R.B.L. (rbliu@phy.cuhk.edu.hk) or J.D. (djf@ustc.edu.cn).

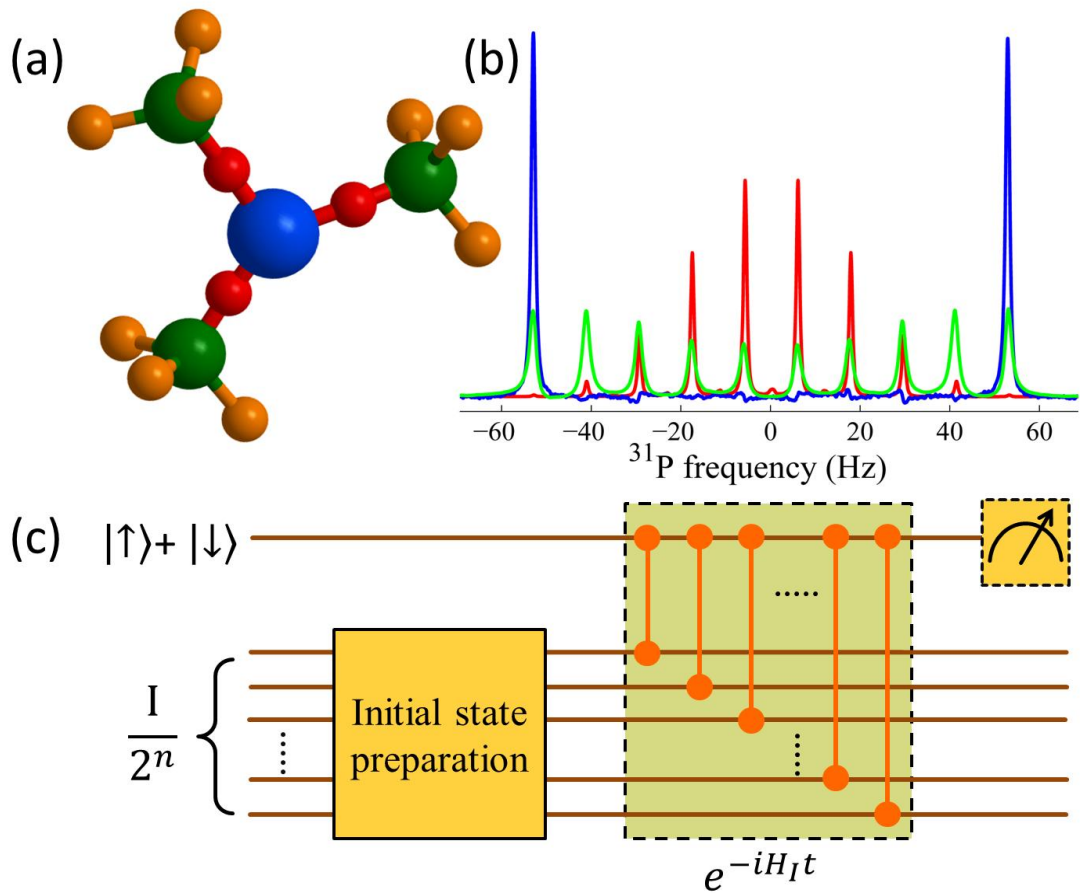

**Figure 1 | System and methods for observation of Lee-Yang zeros. a**, Schematic structure of a TMP molecule. The molecule consists of one $^{31}$P nuclear spin (the blue ball) as the probe and nine equivalent $^1$H spins (the orange balls) as the bath. **b**, Liquid-state $^{31}$P NMR spectrum of TMP molecules at $T=300$ K or $T_{\text{eff}}=\infty$ (red), for the nine $^1$H spins at a simulated temperature $T_{\text{eff}}=15J/8$ (green) and $T_{\text{eff}}=9J/40$ (blue), where $J=2\pi\times 16.75$ sec$^{-1}$ is the coupling between the $^1$H spins. The coupling ($\lambda=2\pi\times 10.57$ sec$^{-1}$) between the local $^1$H spins and the $^{31}$P nuclear spin shifts the resonance frequency of the $^{31}$P by $(9/2-n)\lambda/(2\pi)$, where $n$ is the number of $^1$H spins with $s_j^z=-1/2$. **c**, Quantum circuit for measuring the $^{31}$P spin coherence $L(t)$, with red lines representing the interaction between the probe spin and the bath spins.

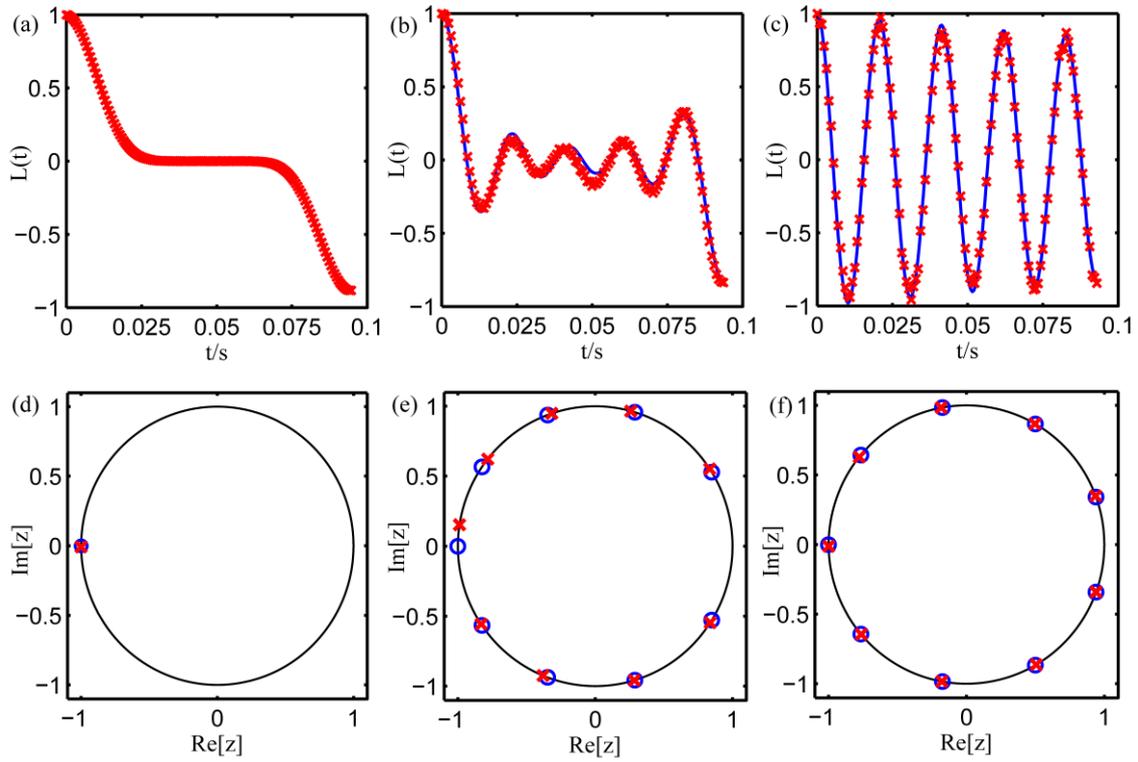

**Figure 2 | Coherence of the $^{31}$P probe spin and the Lee-Yang Zeros.** The effective magnetic field was $h=0$. **a**, **b** & **c** are the measured probe spin coherence $L(t)$ (red symbols) as functions of time for (**a**) laboratory temperature ($T=300$ K), (**b**) simulated temperature $T_{\text{eff}}=15J/8$ and (**c**) simulated temperature $T_{\text{eff}}=9J/40$. The solid lines are the numerically calculated probe spin coherence. **d**, **e** & **f** show the Lee-Yang zeros (by red crosses) measured from the zeros of probe spin coherence corresponding to **a**, **b** & **c**. The theoretical predictions of the Lee-Yang zeros are shown as blue circles for comparison. The unit circles are plotted as a guide to the eye.

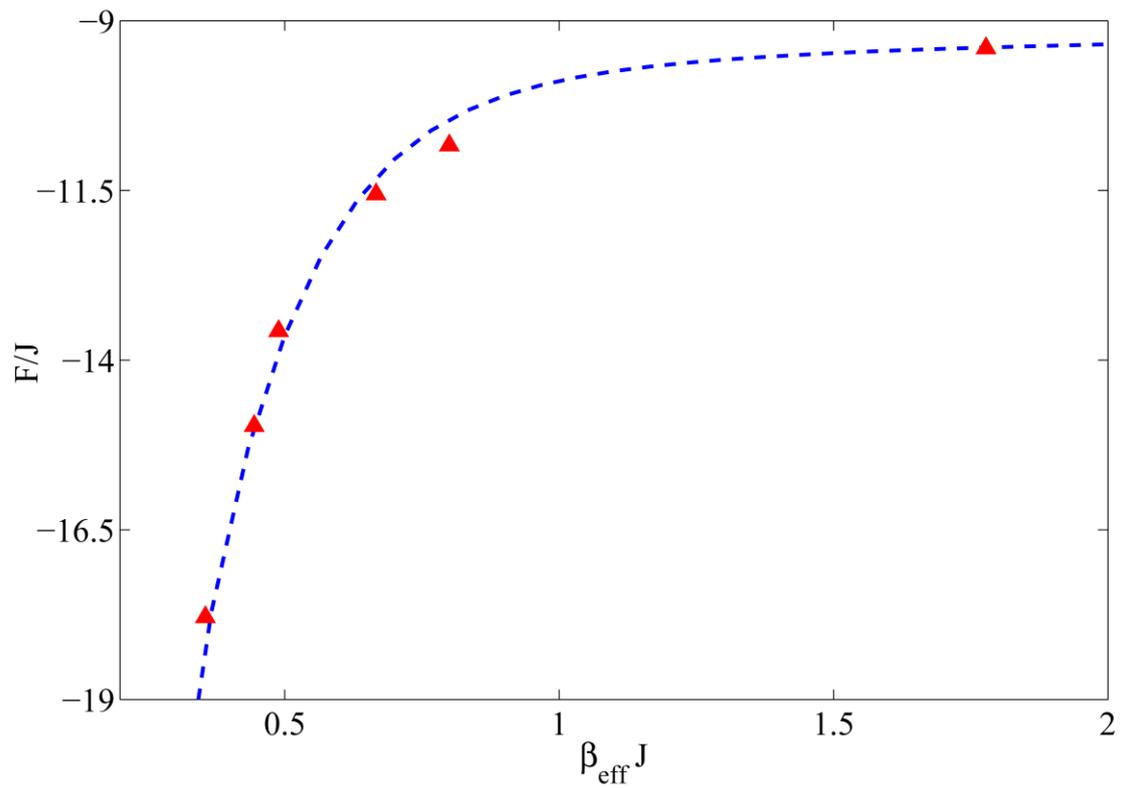

**Figure 3 | Free energy of the $^1$H spin bath reconstructed from the measured Lee-Yang zeros.** The red symbols are the experimentally determined free energy as a function of the simulated inverse temperature. The dashed line is the theoretical calculation.

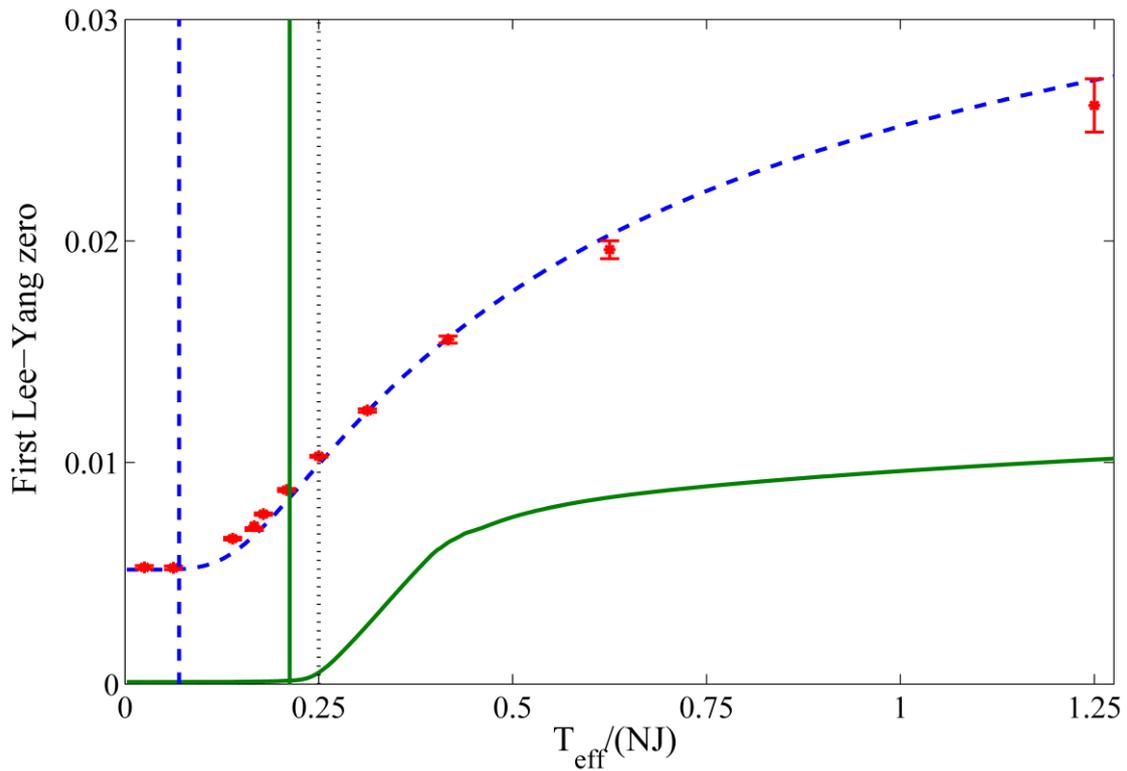

**Figure 4 | Phase transition of the long-range coupling Ising model determined by measurement of Lee-Yang zeros.** The red symbols are the first Lee-Yang zero, i.e., the Yang-Lee singularity edge (given by the first time $t_1$ when the probe spin coherence became zero) as a function of the simulated temperature. The blue dashed line is the numerically calculated result for comparison. The green solid line represents the theoretical curve for a large bath size ($N = 500$) that approximates the thermodynamic limit. Note that the phase transition temperature for the long-range coupling system is proportional to the number of spins *N* (so the *x*-axis is scaled by *N*). Below the critical temperature, the first Lee-Yang zero is almost constant as a function of temperature. The measured critical temperature (indicated by vertical dashed line) deviates from the theoretical predictions for $N = 500$ (the vertical solid line) and for $N \to \infty$ (the vertical dotted line) due to the finite size effect.

# Supplementary Information for "Observation of Lee-Yang zeros"

Xinhua Peng[1], Hui Zhou[1], Bo-Bo Wei[2], Jiangyu Cui[1], Jiangfeng Du[1] & Ren-Bao Liu[2]

1. *Hefei National Laboratory for Physical Sciences at Microscale, Department of Modern Physics, and Synergetic Innovation Center of Quantum Information & Quantum Physics, University of Science and Technology of China, Hefei, 230026, China*

2. *Department of Physics, Centre for Quantum Coherence, and Institute of Theoretical Physics, The Chinese University of Hong Kong, Shatin, New Territories, China*

## S1. Supplementary Equations

### I. The partition function of the long-range Ising model

The Hamiltonian of the long-range Ising model under an external magnetic field is

$$H(h) = -J \sum_{1 \leq i < j \leq N} s_i^z s_j^z - h \sum_{j=1}^{N} s_j^z, \tag{S3}$$

where $s_j^z = \pm 1/2$. The long-range Ising model can be solved by mapping to a large spin. With $S_z \equiv \sum_{j=1}^{N} s_j^z$, the Hamiltonian can be rewritten as

$$H(S) = -\frac{J}{2} S_z^2 - h S_z + \frac{NJ}{8}, \tag{S4}$$

where $S = 0, 1, 2, \cdots, N/2$ for even $N$ and $S = 1/2, 3/2, \cdots, N/2$ for odd $N$. The degeneracy of the large spin subspace is

$$D(S) = C_N^{N/2-S} - C_N^{N/2-S-1}, \tag{S5}$$

where $C_N^n$ is the binomial coefficient. The partition function of the long-range Ising model is

$$\Xi(\beta,h) = \text{Tr}\left[\exp(-\beta H)\right]$$
$$= \sum_S D(S) \sum_{m=-S}^{S} \exp\left(\beta J m^2/2 + \beta hm - N\beta J/8\right) \quad \text{(S6)}$$
$$= e^{N(N-1)\beta J/8} e^{N\beta h/2} \sum_{n=0}^{N} C_N^n e^{\beta J(n^2-Nn)/2} z^n,$$

where $z = \exp(-\beta h)$. The partition function is written as an $N$-th order polynomial of $z$ and its Lee-Yang zeros can be readily evaluated numerically. The probe spin decoherence is related to the partition function of the bath [13]

$$L(t) = \frac{\Xi(\beta, h+i\lambda t/\beta)}{\Xi(\beta,h)}. \quad \text{(S7)}$$

Therefore the coherence of the probe spin coupled to the long-range Ising model is

$$L(t) = \frac{e^{iN\lambda t/2} \sum_{n=0}^{N} C_N^n e^{\beta J(n^2-Nn)/2} \left(e^{-i\lambda t}\right)^n}{\sum_{n=0}^{N} C_N^n e^{\beta J(n^2-Nn)/2}}. \quad \text{(S8)}$$

**II. Phase transitions in the long-range Ising model**

The long-range Ising model has a finite-temperature phase transition between paramagnetism and ferromagnetism. In the large $N$ limit the binomial coefficient can be approximated by the Stirling formula

$$C_N^n \approx \exp\left[-N\left((1/2+x)\ln(1/2+x) + (1/2-x)\ln(1/2-x)\right)\right], \quad \text{(S9)}$$

where $x \equiv n/N - 1/2$. The summation in equation (S6) can be approximated as an integration in the large $N$ limit [28], therefore

$$Z(\beta,h) \approx e^{-N\beta J/8} \int_{-\infty}^{\infty} dx\, e^{-N\varphi(x)}, \quad \text{(S10)}$$

where $\varphi(x) = (1/2+x)\ln(1/2+x) + (1/2-x)\ln(1/2-x) - N\beta J x^2/2$. The integration above can be evaluated by the saddle point approximation. It is clear that the saddle points of the $\varphi(x)$ depend on the value of $\beta J$. If $N\beta J/4 < 1$, the saddle point of $\varphi(x)$ appears at $x = 0$. For $N\beta J/4 > 1$, there are two saddle points symmetrically located at $(-1/2, 0)$ and $(0, 1/2)$. Therefore $N\beta J/4 = 1$ is the phase transition point in the long-range Ising model.

## S.2 Experimental simulation of effective Hamiltonians and effective temperatures

We simulated the thermal equilibrium ensembles of the effective ferromagnetic Ising Hamiltonian $H_{\text{eff}}$ at an effective temperature $T_{\text{eff}}$ and prepared the probe spin in a pseudo-pure state.

The thermal equilibrium of the nuclear spins of TMP molecules is described by the density matrix $\rho_{\text{eq}}^{\text{TMP}} = e^{-\beta H_{\text{TMP}}} / \text{Tr}\left(e^{-\beta H_{\text{TMP}}}\right)$. At room temperature (300 K) and high magnetic field ($B_0 = 9.4$ Tesla), $\rho_{\text{eq}}^{\text{TMP}}$ can be approximated as $\rho_{\text{eq}}^{\text{TMP}} \cong 2^{-10} + \varepsilon_P s_0^z + \varepsilon_H \sum_{1 \leq i \leq 9} s_i^z$, where $\varepsilon_P \approx \beta v_P \approx 2.6 \times 10^{-5}$ and $\varepsilon_H \approx \beta v_H \approx 6.4 \times 10^{-5}$ are the polarizations of the $^{31}$P and $^{1}$H nuclear spins at room temperature, respectively [24, 25]. Hereafter, we only consider the traceless part of the density matrix. As the nine $^{1}$H spins in the bath are magnetic equivalent [29], there are only $N + 1 = 10$ resonant lines in the $^{31}$P spectrum which can be assigned to the $^{1}$H states with the different total magnetization $m = \sum_{1 \leq i \leq N} s_i^z$, with $m = -9/2$, $-7/2$, ..., $9/2$ corresponding to $N + 1$ ensembles states $\rho_m = \sum_\alpha |m\rangle_{\alpha\alpha}\langle m|$, where $\alpha$ runs over all the $C_N^{N/2-m}$ states that have $(N/2 + m)$ spins pointing up and the $(N/2 - m)$ spins pointing down. The degeneracy of each polarization state determines the resonance peak strengths of $^{31}$P at room temperature (see Fig. 1b in the main text).

For the effective ferromagnetic Ising type Hamiltonian $H_{\text{eff}} = -J \sum_{1 \leq i < j \leq 9} s_i^z s_j^z - h \sum_{1 \leq i \leq 9} s_i^z$, the total magnetization $\sum_{1 \leq i \leq 9} s_i^z$ is a conserved quantity. The equilibrium state for the spins with the effective Hamiltonian and at effective temperature $T_{\text{eff}}$ can be simulated by weighted mixture of the states $\rho_m$, that is,

$$\rho_{\text{eff}} = \frac{1}{A} \sum_{m=-N/2}^{N/2} c_m(\beta_{\text{eff}}, h) \rho_m, \quad (S11)$$

where the weighting factor $c_m(\beta_{\text{eff}}, h) = \exp(\beta_{\text{eff}} Jm^2/2) z^{-m}$ and the normalization factor $A = \sum_{m=-N/2}^{N/2} C_N^{N/2-m} c_m(\beta_{\text{eff}}, h)$. Note that this density matrix $\rho_{\text{eff}}$ of the simulated ensemble for the effective ferromagnetic Ising type Hamiltonian $H_{\text{eff}}$ commutes with the microscopic Hamiltonian of the molecules $H_{\text{TMP}}$, so the difference between the microscopic and the effective Hamiltonians would not alter the probe spin coherence evolution. Consequently, we prepared the coupled probe-bath $\rho_{\text{in}}^\Delta = s_0^x \otimes \rho_{\text{eff}}$, where the probe spin was prepared in a pseudo-pure state [24]. Since each resonance line in the $^{31}$P spectrum was labelled by the bath ensemble state $\rho_m$, tuning the control pulse resonant with that particular resonance line amounts to measuring the probe spin coherence with the bath in the corresponding ensemble state $\rho_m$. For the specific case of zero effective field ($h = 0$), the effective bath state $\rho_{\text{eff}}$ is symmetric between $m$ and $-m$. Therefore we performed five different experiments, in each of which two symmetric resonance lines with bath polarizations $m$ and $-m$ were simultaneously excited by a Gaussian-shaped pulse with the length of 550 msec [30, 31]. Then the final results were obtained by summing the probe spin coherence measured in the five experiments using the weighting factors $c_m(\beta_{\text{eff}}, h)$.

## S.3 Experimental data analysis

The probe spin coherence $L(t)$ was detected by the free induction decay (FID) of the $^{31}$P spin in NMR [25]. Its coherence zeros $t_n$ were extracted by fitting the experimental data. The corresponding standard deviations $\eta$ of the probe spin coherence were 0.0005, 0.007, and 0.022 correspondingly for $T_{\text{eff}} = \infty$, $15J/8$, and $9J/40$. The Lee-Yang zeros in the unit circle were obtained by $z_n = \exp(-i\lambda t_n)$. Table SI lists the measured Lee-Yang zeros, along with the corresponding experimental uncertainties $|\Delta\theta_n|$. Here $\theta_n = \lambda t_n$ and $|\Delta\theta_n|$ were calculated from the measurement uncertainty of $L(t)$ using $\Delta\theta_n = \eta / \partial_\theta L(\theta, \beta_{\text{eff}})|_{\theta=\theta_n}$. As shown in Fig. S1, at high temperature ($T_{\text{eff}} \to \infty$), the sensitivity of the probe decoherence $\partial_\theta L(\theta, \beta_{\text{eff}}) \to 0$, leading to a large uncertainty $|\Delta\theta_n|$. Considering this, the experimental Lee-Yang zeros $\theta_n^{\exp}$ for $T_{\text{eff}} \to \infty$ were extracted by a mean value $t_n = (t_1 + t_9)/2$, where $t_1$ ($t_9$) is the start (end) point when the coherence is less than the coherence standard error $\eta$. With decreasing the temperature $T_{\text{eff}}$, $\partial_\theta L(\theta, \beta_{\text{eff}})$ presents sinusoidal oscillations, and the Lee-Yang zeros lie on its troughs and crests where $\partial_\theta L(\theta, \beta_{\text{eff}})$ reaches its maximal magnitude. Consequently, at low effective temperatures, the measurement of the Lee-Yang zeros was more robust against the measurement uncertainty of probe spin coherence $\eta$. For $T_{\text{eff}} = 9J/40$, the experimental uncertainty $|\Delta\theta_n|$ was around 0.005 rad and the maximal relative error was about 2%.

**Table SI | Experimentally measured Lee-Yang zeros at different effective temperatures**. $\Delta\theta_n$ is the uncertainties of the measured Lee-Yang zeros as derived from the standard errors of the probe spin coherence $\eta$ using $\Delta\theta_n = \eta / \partial_\theta L(\theta, \beta_{\text{eff}})\big|_{\theta=\theta_n}$ and $\theta_n^{\text{th}}$ ($\theta_n^{\exp}$) stands for the phase angles of the calculated (measured) Lee-Yang zeros. At room temperature ($T_{\text{eff}} \to \infty$), all of the nine Lee-Yang zeros are degenerate at $z_n = -1$ (i.e., $\theta_n^{\text{th}} = \pi$).

|  |  | $n=1$ | $n=2$ | $n=3$ | $n=4$ | $n=5$ | $n=6$ | $n=7$ | $n=8$ | $n=9$ |
|---|---|---|---|---|---|---|---|---|---|---|
| $T_{\text{eff}} = \infty$ | $\theta_n^{\text{th}}$ | $\pi$ | | | | | | | | |
| | $\theta_n^{\exp}$ | 3.2 | | | | | | | | |
| | $|\Delta\theta_n|$ | 0.9 | | | | | | | | |
| $T_{\text{eff}} = 15J/8$ | $\theta_n^{\text{th}}$ | 0.553 | 1.277 | 1.925 | 2.540 | 3.142 | 3.743 | 4.359 | 5.001 | 5.726 |
| | $\theta_n^{\exp}$ | 0.584 | 1.309 | 1.893 | 2.47 | 2.99 | 3.73 | 4.32 | 5.001 | 5.706 |
| | $|\Delta\theta_n|$ | 0.004 | 0.007 | 0.009 | 0.01 | 0.02 | 0.01 | 0.01 | 0.006 | 0.003 |
| $T_{\text{eff}} = 9J/40$ | $\theta_n^{\text{th}}$ | 0.349 | 1.047 | 1.746 | 2.444 | 3.142 | 3.840 | 4.538 | 5.236 | 5.934 |
| | $\theta_n^{\exp}$ | 0.358 | 1.051 | 1.756 | 2.461 | 3.154 | 3.843 | 4.533 | 5.243 | 5.936 |
| | $|\Delta\theta_n|$ | 0.005 | 0.005 | 0.005 | 0.005 | 0.005 | 0.005 | 0.005 | 0.005 | 0.005 |

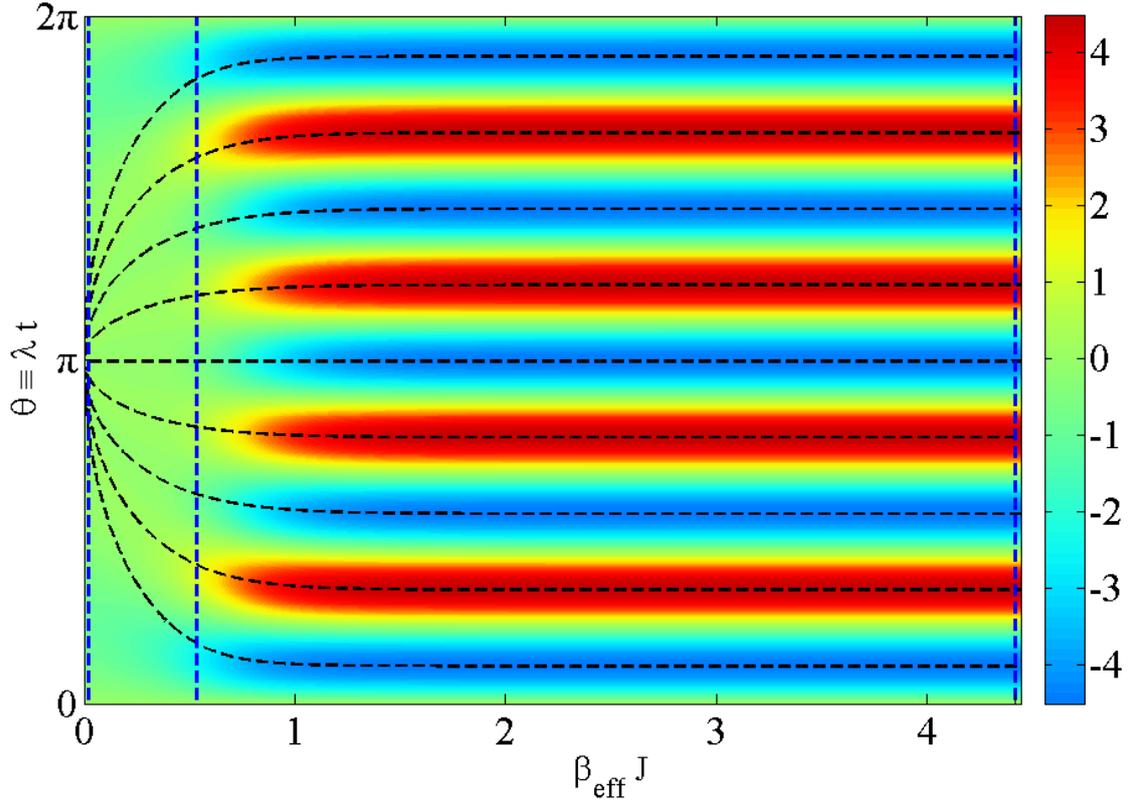

**Figure S1 | Sensitivity of the probe spin coherence to the time variation** $\partial_\theta L(\theta, \beta_{\text{eff}})$. The black dashed lines are the theoretically predicted Lee-Yang zeros as functions of the inverse effective temperature $\beta_{\text{eff}}$. The blue vertical dashed lines indicate the effective temperatures $T_{\text{eff}} = \infty, 15J/8, 9J/40$.

The main error of the probe spin coherence $\Delta L^{\text{exp}}$ came from the imperfection of preparing the initial states. The imperfect initial state in the experiment can be described by

$$\rho_{\text{in}}^{\text{exp}} = \rho_{\text{in}}^{\Delta} + \Delta\rho_{\text{in}}, \tag{S12}$$

where the ideal initial state is

$$\rho_{\text{in}}^{\Delta} = s_0^x \otimes \rho_{\text{eff}} = s_0^x \otimes \frac{1}{A} \sum_{m=-N/2}^{N/2} c_m(\beta_{\text{eff}}, h) \rho_m, \tag{S13}$$

and the experimental deviation can be expressed as

$$\Delta\rho_{\text{in}} = \sum_{m=-N/2}^{N/2} \left( \delta_m^x s_0^x + \delta_m^y s_0^y \right) \otimes \rho_m , \quad (S14)$$

with $\delta_m^x$ and $\delta_m^y$ denoting the deviations on the *x* and *y* components, respectively. The *z* component deviation does not change the probe spin decoherence. This leads to the measured probe spin coherence $L^{\exp}(t) = L(t) + \Delta L^{\exp}(t)$, where the ideal coherence is

$$L(t) = \frac{1}{A} \sum_{m=-N/2}^{N/2} C_N^{N/2-m} c_m \left( \beta_{\text{eff}}, h \right) e^{-im\lambda t} , \quad (S15)$$

and the deviation

$$\Delta L^{\exp}(t) = \sum_{m=-N/2}^{N/2} \left( \delta_m^x - i\delta_m^y \right) e^{-im\lambda t} . \quad (S16)$$

Thus there exist experimental errors in both amplitudes and phases of $L(t)$, and $\left| \Delta L^{\exp} \right| \approx \sum_{m=-N/2}^{N/2} \left| \delta_m^x - i\delta_m^y \right|$. Considering the small experimental errors $\delta_m^x$ and $\delta_m^y$, and the relaxation effect in the experiments, we fitted the experimental data via the real part of the function $\exp(-t/T_2^*) L^{\exp}(t)$ in the parameter range $\delta_m^x, \delta_m^y \in [-0.05, 0.05]$ (the imaginary part of the measure probe spin coherence was zero). Hence, we estimated the state preparation errors $\delta_m^x$ and $\delta_m^y$. The results are listed in Table SII. The maximal error was less than 4%. The experimental deviation $\Delta\rho_{\text{in}}$ was the smallest at $T_{\text{eff}} \to \infty$ and the largest at the intermediate temperature $T_{\text{eff}} = 15J/8$. From these data, we determined that the state fidelities between $\rho_{in}^{\exp}$ and $\rho_{in}^{\Delta}$ were around 0.99, which is consistent with the results of quantum state tomography.

**Table SII | Imperfections of the prepared initial states at various effective temperatures.** The fitted parameters $T_2^*$ are 650 msec, 500 msec and 500 msec, correspondingly for $T_{\text{eff}} = \infty$, $15J/8$, and $9J/40$.

|  | $T_{\text{eff}} = \infty$ | | $T_{\text{eff}} = 15J/8$ | | $T_{\text{eff}} = 9J/40$ | |
|---|---|---|---|---|---|---|
|  | $\delta_m^x$ | $\delta_m^y$ | $\delta_m^x$ | $\delta_m^y$ | $\delta_m^x$ | $\delta_m^y$ |
| $m = -9/2$ | −3.7E−4 | −9.4E−4 | +2.4E−3 | +4.8E−3 | −2.5E−2 | +2.5E−2 |
| $m = -7/2$ | −3.8E−4 | −1.8E−3 | −8.4E−3 | +3.1E−2 | +8.7E−3 | −3.0E−3 |
| $m = -5/2$ | −1.8E−4 | −3.1E−3 | +1.2E−3 | +9.2E−3 | −9.7E−3 | +8.0E−3 |
| $m = -3/2$ | +2.0E−3 | −3.5E−3 | −5.2E−3 | +9.3E−3 | +5.1E−3 | −6.4E−5 |
| $m = -1/2$ | +3.8E−3 | −1.6E−3 | +8.5E−4 | −1.1E−2 | +9.3E−3 | −2.7E−3 |
| $m = +1/2$ | +3.8E−3 | +1.6E−3 | +1.1E−3 | +1.1E−2 | +9.3E−3 | +2.7E−3 |
| $m = +3/2$ | +2.1E−3 | +3.3E−3 | −5.2E−3 | −9.4E−3 | +5.1E−3 | 3.7E−5 |
| $m = +5/2$ | −1.7E−4 | +3.1E−3 | +1.3E−3 | −9.2E−3 | −9.8E−3 | −8.0E−3 |
| $m = +7/2$ | −3.7E−4 | +1.8E−3 | −8.2E−3 | −3.1E−2 | +8.7E−3 | +3.0E−3 |
| $m = +9/2$ | −3.3E−4 | +1.1E−3 | +2.5E−3 | −4.8E−3 | −2.5E−2 | −2.5E−2 |